# Optimizing time-entanglement between two sources.

J.C. Schaake[1], P.G. Evans[2], W.P. Grice[2]

[1]Department of Physics and Astronomy, The University of Tennessee, Knoxville, Tennessee 37996-1200, USA
[2]Quantum Information Science Group, Oak Ridge National Laboratory, Oak Ridge, Tennessee 37831, USA
*Corresponding author: jschaake@vols.utk.edu

**We demonstrate a compact source of four entangled telecommunication wavelength photons, which is used to generate a GHZ state, with minimal spectral and spatial entanglement. The spatial and spectral degree of freedom are minimized by careful source design. To optimize the entanglement between the two sources, distinguishing temporal information must be removed. We demonstrate this high degree of coherence, between pairs of sources, by performing a Hong-Ou-Mandel measurement. This measurement enables the optical path lengths within the system to be equalized, removing timing information from photons. We also measure the second order correlation function to test the rate of multi-pair production from a single pump pulse. With these optimizations completed, we measure a count rate of 13,600 counts/s per mW of pump power.**

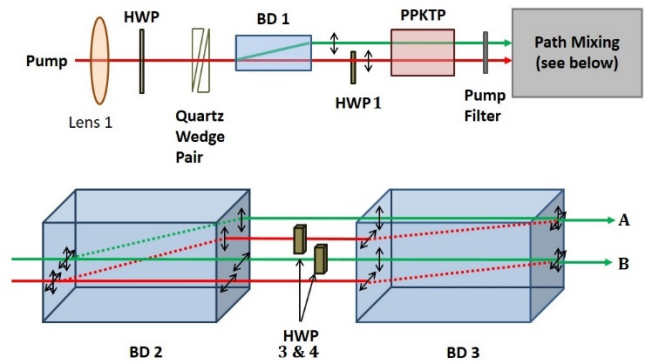

Fig. 1. Schematic of the 2-photon source. Focused pump beam passes through BD1 where it is split into 2 parallel beams. HWP1 ensures proper pump polarization for Type-II SPDC. Emitted photon pairs have paths mixed to generate entangled state.

**Introduction.** Many recent technologies require multi-photon entangled states, often with tailored characteristics. Control of photon wavelength [1]–[4], momentum [5], [6], and entanglement in any degree of freedom are necessary and require high quality photonic state generation. Entangling photons from separate sources requires minimization of spatial and spectral entanglement [7]. After designing and testing a photon-pair source, an alternate configuration is used to generate four-photon states.

The entanglement of photons from separate sources requires the elimination of any identifying information about the photons. Generation of higher photon number states has previously been performed by pumping a single crystal and counting three-photon states at 830 nm at 1/s [8], and four-photon near-IR states with rates of 100-1200/hr [9]–[11]. By pumping multiple identical crystals and heralding, three-photon near-IR states at rates of 30/hr [12], nearly 200/hr [13] and five-photon near-IR states with count rates of nearly 80/hr [14] have been generated. Pumping multiple crystals (or multi-passing a single crystal) has been used to generate even-photon number entangled four near-IR photon states at rates of 15/hr [15], 150/hr [16], six near-IR photon states at a rate of 180/hr [17], eight photon near-IR states at nearly 1/hr [18], and ten photon near-IR states at 75/hr [19].

We have used a combination of the second order correlation function and a Hong-Ou-Mandel [20] measurement to temporally overlap photons from two separate down-conversion sources. The design of this source enables simple scaling to higher photon-number states, limited only by crystal width and one's ability to spatially separate individual beam paths.

We have designed and built an entangled multi-photon source (Fig. 1) with minimal spectral and spatial entanglement [21], which is scalable to higher photon numbers. We characterize the nature of a four-photon version of this source, demonstrating the quality of coherence between our multiple pump beams. The temporal coherence of photons emitted from separate pump beams must be maximized so that the entangled state generated is of the highest purity. We optimize the temporal overlap, which eliminates time-of-flight which-path information, by performing a Hong-Ou-Mandel measurement using pairs of pump beams. By inserting optical delay into the appropriate beam paths, the path lengths, obviously different by simple geometry, can be equalized. When paths are equal, the HOM dip will be at a minimum.

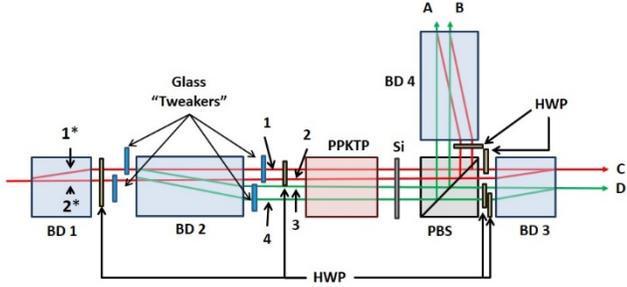

Fig. 2. Schematic of four-photon entangled state generator. Four pump beams are created using a pair of calcite beam displacers (BD1 and BD2). This glass "tweakers" are used to correct any spatial misalignment due to imperfect BD alignment. Four pump beams are incident upon a PPKTP crystal. Down-conversion can take place in any paths. Simultaneous events, in beams 1 and 3 or 2 and 4, lead to emission of a photon at each port A, B, C, and D. PBS separates orthogonally polarized signal and idler photons. Half-waveplates ensure proper polarization for Type-II down-conversion, and path overlap in BD 3 and 4.

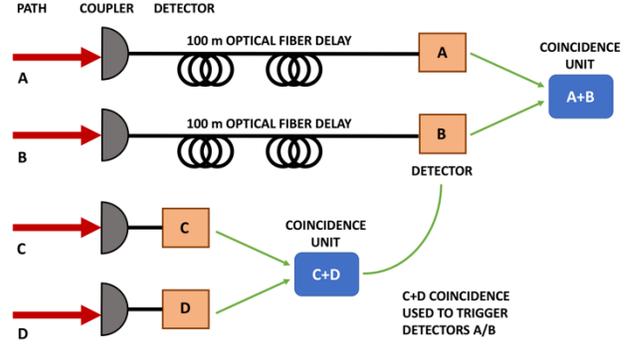

Fig. 3. Schematic of timing. A coincidence, in detectors C and D, is used to trigger detectors A and B. The additional time required for the detection of C/D, the coincidence registration, and the generation of a trigger pulse out must be balanced by the additional travel time of the photons in paths A and B. A 100 m spool of single-mode fiber is inserted into each path A and B. This ensures that a detection at C (or D) coincides with the partner photon being detected at A (or B).

The rate of multiple down-conversion events within a single pump pulse was characterized by measuring the second-order correlation function, or $g^{(2)}$. A single pump beam was used to generate down-conversion events.

**Background.** We begin with a photon source designed to have minimal spatial and spectral entanglement [21]. Minimizing spectral and spatial entanglement is accomplished through proper source design, with crystal parameters and proper focusing being of the utmost importance [22]. Figure 1 shows the experimental schematic. Twin pump beams are created by placing a calcite beam displacer (Fig. 1 BD1) before the pump. This splits the pump beam into two parallel beams with orthogonal polarization. Inserting a half-wave plate (Fig. 1 HWP1) into the proper pump beam gives the correct polarization for Type-II down-conversion. Spontaneous
parametric down-conversion (SPDC) takes place in periodically-polled potassium titanyl phosphate (PPKTP) crystal manufactured by Raicol. A down-conversion event can take place in either path. The paths are rejoined using a second pair of beam displacers. This path mixing generates the final entangled state. It has been shown that this source [21] does indeed have minimal spectral and spatial entanglement.

This design is readily scalable to higher photon numbers. By adding a second calcite beam displacer (BD) before the PPKTP crystal, and adding appropriately placed half-wave plates, we can now pump with 4 identical pump beams (Fig. 2).

**Experiment.** The pump laser is a Coherent Mira, emitting picosecond pulses at 776 MHz at a wavelength of 776 nm. We are pumping a 20 mm long PPKTP crystal. The crystal is cut and poled for degenerate Type-II down-conversion at 1552 nm.

To generate the added set of beams, a second beam displacer is placed before the PPKTP crystal (Fig.2 BD 1 and 2). Two pairs of glass "tweakers" are placed before the down-conversion crystal due to space constraints. These "tweakers" are used to correct any spatial mode misalignment due to imperfect optical axis alignment present in calcite beam displacers. The position of the tweaker is adjusted until both beam paths spatially overlap at the collection fiber. The PPKTP crystal is specified to create degenerate Type-II down-conversion photon pairs at 1552 nm.

As the signal and idler photons have orthogonal polarizations, they can be separated using a polarizing beam splitter. If we use matching beam displacers to recombine the beam paths at A, B, C, and D, the resulting output is a pair of Bell states,

$$\frac{1}{\sqrt{2}}(|H_1 V_2\rangle + |V_1 H_2\rangle),$$

not a single four-photon entangled state. To generate the four-photon state, beam paths 1 and 2 are combined, as are beams 3 and 4. This is accomplished by beam displacer 3 (Fig. 2 BD3). The second beam displacer (Fig. 2 BD4) used combines paths 1 and 3, along with 2 and 4. Now, a four-fold coincidence at ABCD is a four-photon GHZ state:

$$[\frac{1}{2}(|H_A H_B V_C V_D\rangle + |V_A V_B H_C H_D\rangle)].$$

Each collection port uses a Newport 5-axis stage, and a 20x microscope objective, anti-reflection coated for 1550 nm, collecting into a single-mode fiber.

Photon counting is performed by a pair of IdQuantique Id200 single-photon counting modules (SPCM), along with a Princeton Lightwave (PL) PGA-600 SPCM, and an IBM SPCM. The Id200s have stated detection efficiencies of 10%, a minimum gate window of 2.5 ns, dark count rates of 150 Hz, and a maximum trigger rate of 5 MHz. The PL and its forerunner, the IBM SPCM, have detection efficiencies of 10% also, with minimum gating windows of 1 ns, dark count rates of 100 Hz, and maximum triggering rates of 10 MHz. All four detectors can be externally triggered using the Mira fast-photodoide output.

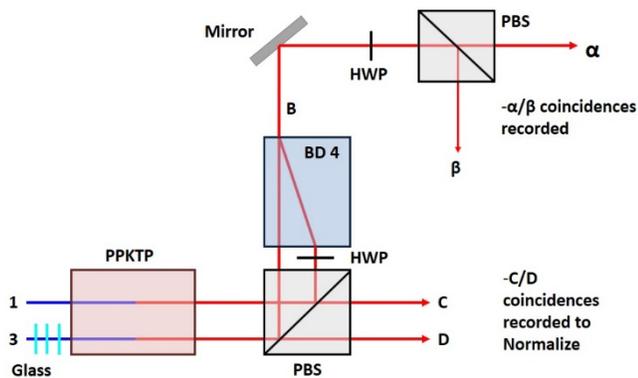

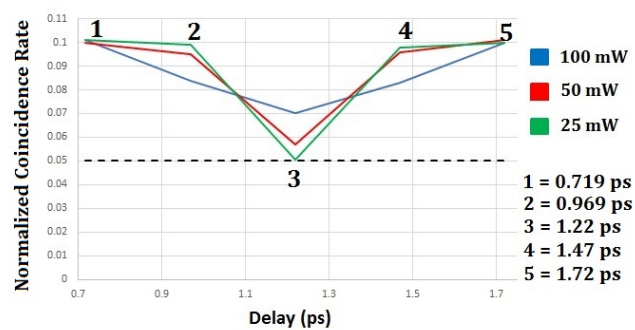

Fig. 4. Schematic of the HOM apparatus. Simultaneous down-conversion events in beams 1 and 3 lead to a signal photon pair at ports C and D. The idler photons are both reflected to BD 4, where their paths are combined and exit at B. A HWP rotates the H and V polarization to ±45°. The photons are split by a PBS and collected at α and β. C/D and α/β coincidence rates, along with α and β singles rates are recorded as the number of cover slips in each beam path is varied.

Fig. 5. Plot of Coincidence rate as a function of delay. The delay steps are individual microscope cover slip thicknesses added to path 3. Plots for three different pump powers are shown. As the pump power is reduced, the probability of generating a second, accidental down-conversion event within the same pump pulse drops as expected. The measured minimum is shown with a dashed line. Actual delays are denoted on the plot by 1, 2, 3, 4, and 5.

To minimize timing calibration issues, photons were collected at ports C and D using the Id200 detectors, as the IBM detector did not have an adjustable internal delay to match the gate to the pulse arrival. The Mira fast-photodoide output was sent to a Quantum Technologies Divide By device. The 76 MHz Mira signal was reduced 16x to trigger the Id200 detectors at 4.76 MHz. The HWP before BD 2 (Fig. 2) was adjusted to balance count rates in both arms.

The NIM output signal from each detector was sent to an ORTEC Quad input 4020 coincidence module. Any coincidence registered was sent to a Stanford Research Systems DG 535 delay module. This signal will function as the trigger for the other two detectors. The DG535 takes an external trigger and can generate two user defined, delayed outputs with step sizes as small as 5 ps.

As we are collecting four photons at four different couplers, the detectors need to be synchronized (Fig. 3). We use a coincidence in ports C and D to herald photons in paths A and B. This coincidence generates a trigger pulse, which is sent to a SRS DG535 delay module. The DG535 allows for fine tuning of the trigger delay. The delayed output pulse is now sent to detectors A and B. To account for the time needed to detect photons, register a coincidence, and send out a trigger pulse, 100 m spools of single-mode optical fiber are placed between the ports A and B and the coinciding detectors. This additional optical path length matches the electronic delay required, enabling proper triggering of detectors A and B.

To match delays for all 4 detectors, all but path 1 is blocked. This only allows for a photon pair to reach ports B and C. The DG535 output channel 1 is scanned until a peak in coincidences is seen. This peak occurs only when both detectors are seeing the same pulse. Path 2 is unblocked and ports C and A are matched in the same manner by scanning the DG535 channel 2 output.

As there are no shared paths for ports C and D, we must match D to either detector A or B to ensure proper timing. Either path 3 or 4 is unblocked and the internal delay adustment on the Id200 is scanned until peak coincidences are found.

If beam paths are traced, it becomes evident that simultaneous down-conversion events in every beam pair do not result in photons arriving in all four exit ports. Only down-conversion events in beams 1 and 3, or in beams 2 and 4, give photons at all four ports.

Optimizing the overlap of the down-conversion events requires eliminating timing information which comes from the optical path length differences generated within the beam displacers. Simple geometric considerations show that the optical path lengths of photons in paths 3 and 4 (Fig. 2) are longer than those in paths 1 and 2. This results in later arrivals for pump pulses in paths 3 and 4 at the PPKTP crystal. In BD 3 (Fig. 2), photons in paths 2 and 4 have longer OPL than those in paths 1 and 3. Finally, in BD 4 (Fig. 2), photons in paths 1 and 2 have longer OPL that those in paths 3 and 4.

To optimally overlap the optical paths, a HOM measurment was performed using photons from pairs of pump beams. This was accomplished by adjusting the pump half-wave plates to only generate pairs of pump beams instead of the normal four beams. By adjusting a HWP just before BD 1 (Fig. 2), we can generate either beam 1*, beam 2*, or both. If we only generate beam 1*, we can have down-conversion events in beams 1 and 3. Simultaneous down-conversion events in beams 1 and 3 generate an orthogonally polarized signal and idler pair in each path. The H-polarized photons are transmitted by the PBS and collected into single-mode fibers at C and D. The V-polarized photons are reflected at the PBS and pass into BD 4 (Fig. 4). BD 4 overlaps paths 1 and 3 into output B.

Output B (Fig. 4) passes through a HWP set to 22.5° to mix the H and V polarixed photons into the ±45° basis (Fig. 4). Now, each photon incident upon the PBS is reflected or transmitted with equal probability. This leads to coincidences at couplers α and β.

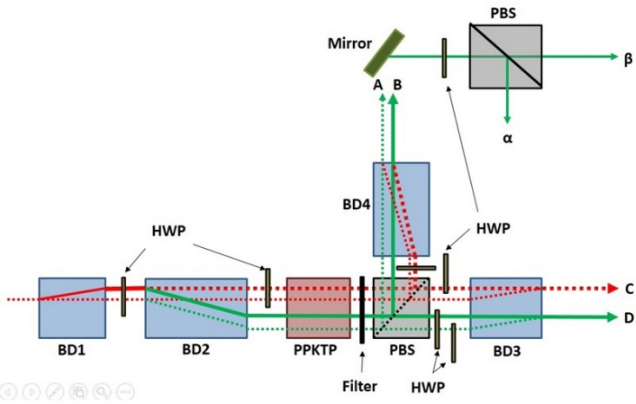

Fig. 6. Schematic of apparatus used to collect second order correlation function data. Only a single pump beam is used, denoted here by a solid line. A detection at D is used to herald the presence on a photon in B. Singles rates at α and β are recorded along with α-β coincidences. Dashed lines represent other paths photons take through system.

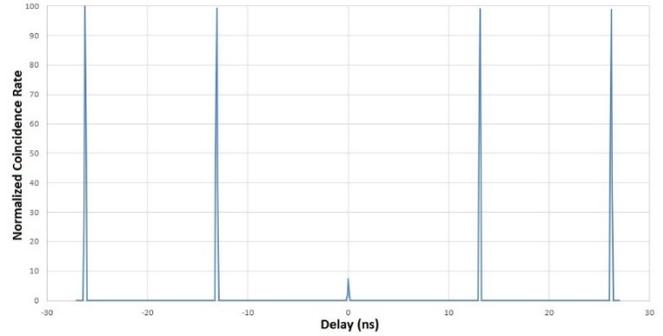

Fig. 7. Coincidence rate versus relative detector delay using a single pump beam. The peaks are due to the pulsed laser source. When each detector gate overlaps with a pulse, coincidences occur at a background rate. 13.1 ns separates the pulses. The zero-delay peak is where the detectors are looking at identical pulses. The matching peak has a minimum of 7.9%, demonstrating that this source reliably produces single-photon pairs.

If photons from different paths do not simultaneously arrive at the beam splitter, the coincidence rate increases.

**Data.** A coincidence at C and D was used to herald the presence of a pair of single-photons in path B. The singles rates at α and β, along with the α/β coincidences were recorded. The singles rates were used to normalize the coincidence rate as the insertion of uncoated glass into the system increased losses.

Collected data for three different pump powers is shown above (Fig. 5). For 100 mW of pump power, a dip to 0.0701 is seen with an optical delay of 1.22 ps. This is below the 0.1 time mismatched rate seen for other delays. As the pump power is reduced, the dip becomes more pronounced. For a 50 mW pump, the minimum drops to 0.0570. Finally, for a 25 mW pump power, the minimum drops to 0.0505. This is well above the theoretical minimum of zero, but the appearance of a dip shows that the pulses are nearly overlapped.

Two likely explanations for not reaching the theoretical minimum are the inablility to rotate the small half-waveplates just before the beam displacers, and the delay step sizes being restricted to individual cover slips. As the wave-plates before the beam splitters are not able to be rotated, there is a reasonable possiblilty of photons not being folded into the correct path. With the minimum delay step being a single cover slip, it is very possible that we are simply missing the true minimum with exact timing matching. A glass wedge pair would give us more adjustibility but not likely be small enough to fit into the necessary spaces with this compact source design.

To ensure that the coincidences measured are from pairs of photon events and not higher photon number causes, we perform a Hanbury-Brown Twiss [23] measurement on our source.

Only a single pump beam is used here as we are measuring photon anti-bunching, or the rate that the source produces multiple down-conversion events within a single pump pulse.

The pump power was set to 25 mW, and only a single pump beam (path 3) was passed through the system (Fig. 6) This should result in a photon in paths D and B. A detection at port D is used to herald detectors α and β. The 100m SMF spools are left in place to keep timing settings the same. The delay setting for detector α was set to a peak, while the delay for detector β was scanned across 60 ns.

As seen in the plot of coincidence rate as a function of delay (Fig. 7), the coincidence rate drops to 7.9% of the time mismatched rate. This shows that our source does indeed generate single-photon pairs consistently. The remaining accidentals are likely a function of imperfect polarization rotation by the wave-plates. The small size of the wave-plates and the compact design of the experiment does not allow for fine tuning of the polarizer angle. Also, greater reduction of the pump power would further suppress the multi-pair production rate. This was not done as it greatly increases integration time for coincidence counting.

With the timing optimization performed, we measured the four-photon state rate of our source. We measured 8 four-photon events per minute at the lowest power (25 mW). Using the stated detector efficiencies of 10%, pump power (25 mW), and the repetition rate of the laser (76 MHz), this extrapolates to an estimated four-photon rate of 13,600 events/s per mW of pump power. These photons have minimal spectral, spatial, and time entanglement. A full state tomography was not performed due to time constraints.

**Conclusion.** We have demonstrated a four-photon source that entangles photons from a pair of two-photon sources. We demonstrate the coherence of the sources by making a Hong-Ou-Mandel measurement. The minimum of this measurement was 0.0505, which demonstrates that the pair of pump beams have been optimally temporally overlapped. This method is a useful tool that can be used to remove timing information when entangling photons from separate sources.

The second order correlation function was measured for a single pump beam. The coincidence rate drops to 7.9% of the background rate when the detector pair experience zero relative delay. This demonstrates that at a pump power of 25 mW, this source reliably produces single pairs of down-converted photons within a single pump pulse.

With timing information reduced, a production rate of 13,600 four-photon events per second, per mW of pump power were measured. These telecommunication wavelength photons have minimal spatial, spectral, and time entanglement.